\documentclass[3p,twocolumn]{elsarticle}
\biboptions{sort&compress}

\usepackage[utf8]{inputenc}
\usepackage[T1]{fontenc}
\usepackage{lmodern}
\usepackage{microtype}
\usepackage{graphicx}
\usepackage{booktabs}
\usepackage{amsmath}
\usepackage{siunitx}
\DeclareSIUnit\angstrom{\text{\AA}}
\DeclareSIUnit\bohr{\mathit{a}_{0}}
\DeclareSIUnit\elementarycharge{\mathit{e}}
\usepackage{stfloats}
\usepackage{xurl}
\usepackage[colorlinks=true,allcolors=blue]{hyperref}
\usepackage{orcidlink}
\AtBeginDocument{}

\journal{Materials Today Physics}

\begin{document}

\begin{frontmatter}

\title{Facet- and thickness-dependent band-edge alignment at ZrSe\texorpdfstring{$_3$}{3} surfaces: hybrid-functional calculations with spin-orbit coupling}

\author[1,2]{Sandro G. Holanda\texorpdfstring{\,\orcidlink{0000-0003-3367-9729}\corref{}}{}}
\author[3]{Bruno Ipaves\texorpdfstring{\,\orcidlink{0000-0003-2908-4589}\corref{cor1}}{}}
\ead{ipaves@unicamp.br}
\author[4]{Astrid Campos-Mata\texorpdfstring{\,\orcidlink{0000-0002-2138-562X}\corref{}}{}}
\author[4]{Shreyasi Chattopadhyay\texorpdfstring{\,\orcidlink{0000-0003-4429-6117}\corref{}}{}}
\author[4]{\texorpdfstring{\\}{}Pulickel M. Ajayan\texorpdfstring{\,\orcidlink{0000-0001-8323-7860}\corref{}}{}}
\author[3]{Douglas S. Galv\~{a}o\texorpdfstring{\,\orcidlink{0000-0003-0145-8358}\corref{cor1}}{}}
\ead{galvao@ifi.unicamp.br}
\author[1,2,4,5]{Marcelo L. Pereira Junior\texorpdfstring{\,\orcidlink{0000-0001-9058-510X}\corref{cor1}}{}}
\ead{marcelo.lopes@unb.br}
\cortext[cor1]{Corresponding author.}

\affiliation[1]{organization={Materials Science Postgraduate Program, College UnB Planaltina, University of Bras\'{i}lia}, city={Bras\'{i}lia}, postcode={73380-900}, state={DF}, country={Brazil}}
\affiliation[2]{organization={NanoEngineering Laboratory, University of Bras\'{i}lia}, city={Bras\'{i}lia}, postcode={70910-900}, state={DF}, country={Brazil}}
\affiliation[3]{organization={Applied Physics Department and Center for Computational Engineering and Sciences, State University of Campinas}, city={Campinas}, postcode={13083-970}, state={SP}, country={Brazil}}
\affiliation[4]{organization={Department of Materials Science and Nanoengineering, Rice University}, city={Houston}, postcode={77005}, state={TX}, country={USA}}
\affiliation[5]{organization={College of Technology, University of Bras\'{i}lia}, city={Bras\'{i}lia}, postcode={70910-900}, state={DF}, country={Brazil}}

\begin{abstract}
Transition-metal trichalcogenides MX$_3$ are quasi-one-dimensional van der Waals materials whose layers form through lateral chain binding, giving them an interlayer cleavage plane and in-plane electronic anisotropy. Among them, ZrSe$_3$ is stable under ambient conditions and semiconducting. Exfoliated crystals expose four crystallographic planes, and electrocatalytic measurements on an individual crystal assigned the hydrogen evolution activity to the high-energy (210) facet rather than to the basal plane. Those non-basal surfaces remain uncharacterized, since the available calculations treat either the bulk crystal or the free-standing basal monolayer. Here, we compute the surface energies and the equilibrium morphology of these four surfaces with a dispersion-corrected semilocal functional, and their band-edge alignment with the HSE06 hybrid functional including spin-orbit coupling. The surface energies span nearly an order of magnitude, from \SI{0.097}{\joule\per\meter\squared} for (001) to \SI{0.869}{\joule\per\meter\squared} for (010), and the Wulff shape exposes the four observed planes with area fractions of \SI{17.1}{\percent}, \SI{4.4}{\percent}, \SI{71.4}{\percent}, and \SI{7.0}{\percent}. The ionization potential varies by \SI{0.45}{\electronvolt} between facets and the electron affinity by \SI{0.80}{\electronvolt}, the gap of (210) falls from 1.43 to \SI{0.94}{\electronvolt} between 10 and \SI{23}{\angstrom}, and (010) remains metallic. The facets separate according to whether the cut preserves or distorts the Se-Se dimer, and the distorted dimers concentrate the frontier states. Because the same dimer governs the bulk electronic structure of the MX$_3$ family, the criterion transfers to the related trichalcogenides.
\end{abstract}


\end{frontmatter}

\section{Introduction}

Transition-metal trichalcogenides MX$_3$, with M a group IV or V transition metal and X a chalcogen, are quasi-one-dimensional van der Waals materials. This name reflects a structural hierarchy uncommon among layered solids, since their layers are built from one-dimensional chains rather than from a two-dimensional network \cite{Chen2023, Patra2020, Island2017}. Each chain is a stack of trigonal prisms sharing triangular faces, with the metal atom at the center of six chalcogens, and all chains in a crystal run parallel to one another. The chains bind laterally into layers, and the layers stack through van der Waals forces, so the crystals combine a two-dimensional cleavage plane with a one-dimensional electronic anisotropy. Work on two-dimensional metal chalcogenides more broadly has shown that monolayers can be grown directly by solid-vapor reaction \cite{Li2016} and that their surfaces can be functionalized through Lewis acid-base chemistry \cite{Lei2016}, so the exposed face is a handle on the material and not only its boundary. Among chalcogenides, ZrSe$_3$ has attracted particular attention because it is stable under ambient conditions and semiconducting, and it has been used in visible-light detectors \cite{Xiong2015}, polarization-sensitive photodetectors \cite{AlcazarRuano2025}, thermoelectric devices \cite{Zhou2018}, and thin-film transistors \cite{Thole2022}.

What that hierarchy does electronically follows from a second inequivalence, this time among the chalcogen atoms of a single prism. Two of the three occupy a dimer, so the formal electron count reads M$^{4+}$(X$_2$)$^{2-}$X$^{2-}$, and whether the compound is a semiconductor or a metal follows from where the X-X antibonding $\sigma^{*}$ level sits relative to the Fermi energy \cite{Jellinek1974, Bullett1979, CanadellWhangbo1991}. The argument is quantitative in NbSe$_3$, where three inequivalent chains are distinguished by Se-Se separations of 2.37, 2.49, and \SI{2.91}{\angstrom}, and the two more weakly paired chains carry the metallic bands \cite{Canadell1990, Hodeau1978}. The same lever operates under pressure in TiS$_3$, where the S-S pair lengthens until the primary and secondary bonds interchange and the compound metallizes \cite{AbdelHafiez2024}, and under chemical reduction of ZrSe$_3$, where the added electrons populate the Se-Se $\sigma^{*}$ and lengthen the diselenide pair \cite{Elgaml2022}. The dimer is therefore the structural variable controlling the family's electronic structure.

No calculation has yet tested that argument on a ZrSe$_3$ surface, since work on the compound has consistently addressed either the bulk crystal or the free-standing basal monolayer. Hybrid-functional work places the bulk indirect electronic gap at \SI{0.75}{\electronvolt} and the monolayer between 1.01 and \SI{1.17}{\electronvolt} \cite{Jin2015, Mortazavi2022, LiDaiZeng2015}. The $GW$ approximation, a many-body correction to the quasiparticle energies, predicts \SI{0.66}{\electronvolt} for the bulk and \SI{1.63}{\electronvolt} for the monolayer \cite{Li2022, Zhou2018}, and optical absorption on single crystals yields an indirect gap of \SI{1.1}{\electronvolt} and a direct gap of \SI{1.47}{\electronvolt} \cite{Patel2005}, with a direct excitonic gap at \SI{1.818}{\electronvolt} at low temperature \cite{Kurita1993}. Transport measurements on exfoliated material give a gap of \SI{0.6}{\electronvolt} that widens as the samples thin \cite{Thole2022, Hollmann2025}. In every case, the direction being cut is the one perpendicular to the layers, which is also the direction along which the crystal cleaves on its own. The free-standing monolayer is not a surface calculation, since it carries no information on the structural edges (terminations) or surface energy, and consequently neither the relative stability of the possible ZrSe$_3$ terminations nor the equilibrium shape of the crystal has been established.

This information is important because exfoliated ZrSe$_3$ does not expose only its cleavage plane. In an earlier work \cite{CamposMata2026}, we identified by high-resolution transmission electron microscopy and selected-area electron diffraction four crystallographic planes in single-crystal ZrSe$_3$ ribbons, (100), (010), (001), and (210). Also, using an electrocatalytic microcell on an individual crystal, we measured a hydrogen evolution overpotential of \SI{431}{\milli\volt} at \SI{10}{\milli\ampere\per\centi\meter\squared} with a Tafel slope of \SI{82.3}{\milli\volt} per decade. We assigned this activity to the high-energy (210) facet rather than to the conventionally expected basal plane. The active plane is therefore one of those left out of the available electronic description, and the quantity that would determine how much of it a real crystal exposes, the equilibrium morphology, remains an open question.

In this work, we computed the surface energies, equilibrium morphology, and electronic band-edge alignment of the (100), (010), (001), and (210) ZrSe$_3$ surfaces. We obtained geometries and total energies using the Perdew, Burke, and Ernzerhof (PBE) functional with the nonlocal many-body dispersion model (MBD-NL). The electronic structure was obtained from the hybrid functional of Heyd, Scuseria, and Ernzerhof (HSE06) \cite{Heyd2003, Heyd2006, Krukau2006} with spin-orbit coupling (SOC). Surface energies determine the equilibrium shape of the crystal and the relative areas of its four facets. At the same time, the electronic band structures allow us to estimate the ionization potential (IP) values, electron affinity (EA), and effective masses of each facet on an absolute scale, along with their evolution with crystal thickness. The four facets are distinguished by whether the cut preserves or distorts the Se-Se dimer, which connects the surface behavior of ZrSe$_3$ to the bonding argument that organizes the bulk electronic structure of the MX$_3$ family. Figure~\ref{fig:sistema} shows the bulk crystal and the four terminations that this analysis addresses.

\section{Computational details}

All calculations were performed with the all-electron, full-potential code FHI-aims \cite{Blum2009}, which expands the Kohn-Sham states in numeric atom-centered orbitals. The tight defaults of the 2020 species set were used for Zr and Se, scalar-relativistic effects were treated in the atomic zeroth-order regular approximation, and the occupations were broadened with a Gaussian of \SI{0.01}{\electronvolt}. No calculation was spin polarized. That choice was tested on the one surface that comes out metallic, where a fixed-moment scan places every polarized solution within \SI{6}{\milli\electronvolt} of the unpolarized one up to a moment of $0.5\,\mu_{\mathrm{B}}$ per cell and above it thereafter. Self-consistency was converged to \SI{e-6}{\elementarycharge\per\bohr\cubed} in the density, \SI{e-3}{\electronvolt} in the sum of eigenvalues, and \SI{e-6}{\electronvolt} in the total energy.

We relaxed the geometries with the PBE exchange-correlation functional combined with MBD-NL \cite{Hermann2020}. The bulk cell was relaxed in both the lattice vectors and the internal coordinates on a $10\times14\times6$ $k$-point grid until the maximum residual force fell below \SI{e-3}{\electronvolt\per\angstrom}.

Slabs were cut from the relaxed bulk along (100), (010), (001), and (210), the four planes that we resolved experimentally \cite{CamposMata2026}, with thicknesses of $N$ structural units and at least \SI{24.8}{\angstrom} of vacuum. The in-plane lattice vectors were fixed at the bulk values, and all atomic positions were relaxed with the same force criterion. Every slab is stoichiometric and terminates in two equivalent faces, so its surface energy carries no dependence on the selenium chemical potential and no dipole correction is needed. We chose in-plane $k$-point grids per facet to maintain comparable sampling density: $5\times13\times1$ for (100), $9\times5\times1$ for (010), $9\times13\times1$ for (001), and $5\times5\times1$ for (210).

Electronic structures were obtained with the HSE06 screened hybrid functional, using a screening parameter of \SI{0.2}{\per\angstrom} and \SI{25}{\percent} exact exchange, on the relaxed geometries, keeping the in-plane grids of the relaxations for the slabs and reducing the bulk grid to $8\times12\times5$. SOC was included non-self-consistently, a treatment that reproduces the self-consistent result for elements up to $Z \approx 50$ \cite{HuhnBlum2017}, which covers Zr and Se, and it narrows the gaps reported below by \num{0.02} to \SI{0.06}{\electronvolt}. Band gaps were taken as the minimum over both the self-consistent $k$-point grid and the band path, since the grid alone overestimates the gap whenever an extremum falls between grid points. Band-edge positions were referenced to the vacuum level obtained from the electrostatic potential in the vacuum region of each slab \cite{VandeWalle1987, Hinuma2014}.

\section{Results and discussion}

\begin{figure*}[tb]
\centering
\includegraphics[width=0.7\linewidth]{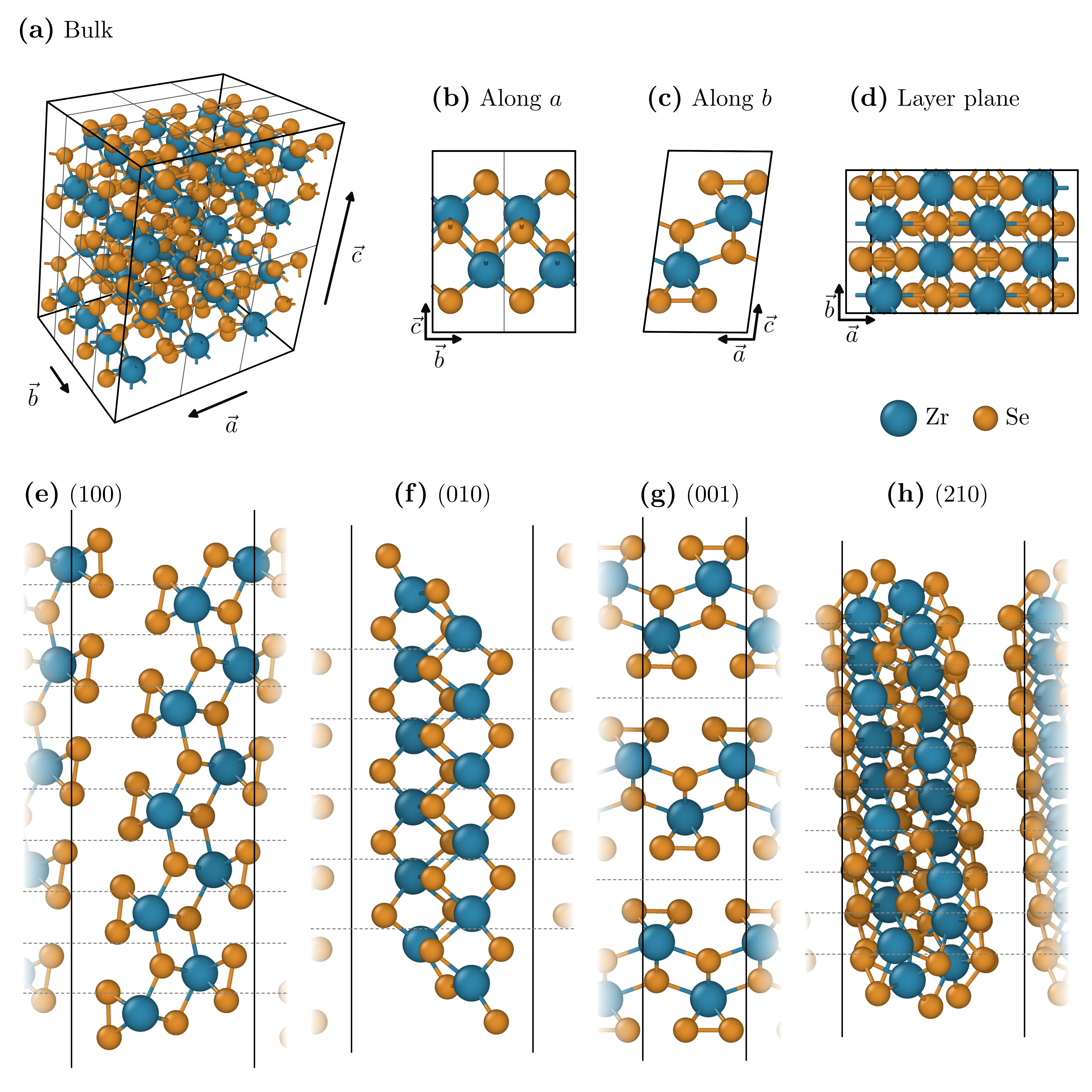}
\caption{Bulk ZrSe$_3$ in perspective (a) and in lateral projections (b--d), and the slabs obtained from the (100), (010), (001), and (210) cuts (e--h). Unit cells are outlined, and the thickness counter $N$ is marked.}
\label{fig:sistema}
\end{figure*}

ZrSe$_3$ crystallizes in the monoclinic space group $P2_1/m$, in which the trigonal-prismatic chains run along the $b$ axis, bind laterally into layers spanned by $a$ and $b$, and stack along $c$, with $\beta$ being the angle between $a$ and $c$, as Figure~\ref{fig:sistema} shows for the bulk structure and the four terminations on a common scale. The chains are seen along $b$ in Figure~\ref{fig:sistema}(a), and the lateral projections in Figures~\ref{fig:sistema}(b)--(d) make the layered stacking explicit. Relaxation gives $a = \SI{5.417}{\angstrom}$, $b = \SI{3.738}{\angstrom}$, $c = \SI{9.567}{\angstrom}$, and $\beta = 97.36^{\circ}$, which reproduces the room-temperature diffraction values of 5.415, 3.753, and \SI{9.473}{\angstrom} with $\beta = 97.72^{\circ}$ to better than \SI{1}{\percent} in every lattice parameter \cite{Xu2022, Furuseth1975}. The four planes resolved by electron microscopy cut that crystal in ways that differ in what they break. The slabs derived from each cut appear in Figures~\ref{fig:sistema}(e)--(h), in the order (100), (010), (001), and (210), with the unit cell and the thickness counter $N$ marked on each. Only the (001) cut separates layers without breaking bonds, so it is the plane along which the crystal exfoliates. The (100) cut runs parallel to the chains and separates them laterally, the (010) cut is perpendicular to the chains and severs them, and the (210) cut crosses both the chain direction and the stacking direction.

\begin{figure*}[tb]
\centering
\includegraphics[width=\linewidth]{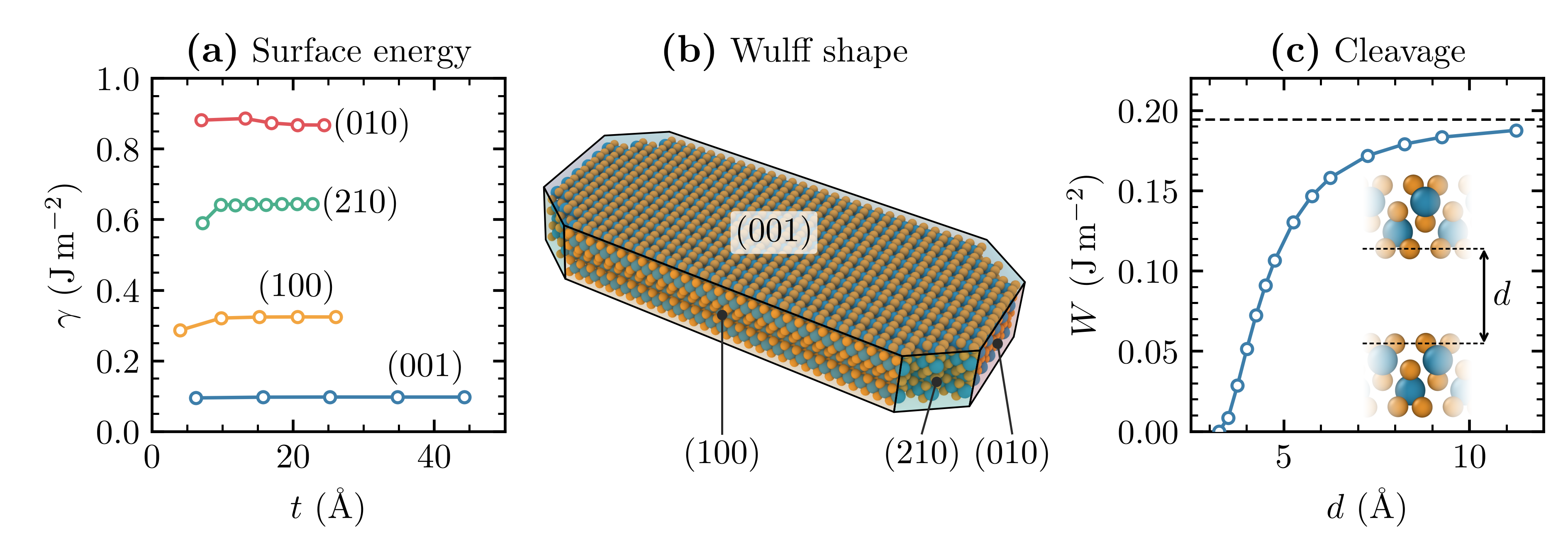}
\caption{Surface energy against slab thickness (a), equilibrium Wulff shape with the facet families labeled (b), and cleavage energy of the (001) plane against the separation $d$, with a schematic of the separated crystal in the inset (c).}
\label{fig:estabilidade}
\end{figure*}

The above-mentioned hierarchy of broken bonds is reflected quantitatively in the surface energies, which span nearly an order of magnitude across the four facets. The excess energy carried by a surface is defined as:

\begin{equation}
\gamma = \frac{E_{\text{slab}} - n E_{\text{bulk}}}{2A},
\label{eq:gamma}
\end{equation}

\noindent where $E_{\text{slab}}$ is the total energy of a slab containing $n$ formula units, $E_{\text{bulk}}$ is the bulk energy per formula unit, $A$ is the area of the in-plane cell, and the factor of two accounts for the two faces the slab exposes. A smaller $\gamma$ means a termination that costs less energy to create, and therefore a more stable surface.

Figure~\ref{fig:estabilidade}(a) follows $\gamma$ along each thickness series, with total energies from PBE and MBD-NL, and the values collected in Table~\ref{tab:estabilidade} average each series over the thicknesses beyond which it stops drifting, three structural units for (001) and four for the other three cuts. They are \SI{0.323}{\joule\per\meter\squared} for (100), \SI{0.869}{\joule\per\meter\squared} for (010), \SI{0.097}{\joule\per\meter\squared} for (001), and \SI{0.643}{\joule\per\meter\squared} for (210). The (100) series is restricted to even $N$, since odd $N$ produces a slab with two inequivalent faces, the (010) cut admits two terminations, of which the one used throughout lies \SI{0.05}{\joule\per\meter\squared} below the other, and the two-unit (210) slab reconstructs and was left out of its series. The ordering follows the coordination that each cut destroys, and the (001) value is smaller than the others by a factor of three to nine because that termination breaks no covalent bonds. The last two thicknesses of every series agree to within \SI{0.001}{\joule\per\meter\squared}, which confirms that the bulk reference in Equation~\ref{eq:gamma} is consistent with the slab calculations \cite{Boettger1994, Fiorentini1996}.

Wulff's theorem turns those four numbers into a shape, with the distance from the crystal center to each facet proportional to its surface energy \cite{Wulff1901, Herring1951}. Evaluating it for the point group $2/m$ over the four families gives the polyhedron shown in Figure~\ref{fig:estabilidade}(b), on which they take area fractions of \SI{17.1}{\percent}, \SI{4.4}{\percent}, \SI{71.4}{\percent}, and \SI{7.0}{\percent} for (100), (010), (001), and (210). It ranks the planes that diffraction resolved \cite{CamposMata2026} against one another and not against all possible orientations, and the fractions are the weights to use when a property is averaged over the area a crystal exposes. The elongated habit of the exfoliated crystals is not reproduced by the equilibrium shape, which is expected for a morphology set by growth kinetics and by mechanical exfoliation along the chains rather than by surface thermodynamics.

Because the basal plane dominates that shape, its surface energy is the one worth reaching by a second route. We open a fracture between two layers of a basal slab, hold the atoms fixed, and follow the energy cost per unit area as a function of the separation $d$ between the two halves, whose derivative is the stress that the interlayer bonding sustains:

\begin{equation}
W(d) = \frac{E(d) - E(0)}{A}, \qquad \sigma(d) = \frac{\mathrm{d}W}{\mathrm{d}d}.
\label{eq:cleavage}
\end{equation}

\noindent Figure~\ref{fig:estabilidade}(c) shows $W(d)$, which reaches \SI{0.188}{\joule\per\meter\squared} at a separation of \SI{8}{\angstrom} and is still increasing along the van der Waals tail, within \SI{4}{\percent} of the value $2\gamma_{(001)} = \SI{0.194}{\joule\per\meter\squared}$ expected from Equation~\ref{eq:gamma}. The two routes agree, although one holds the fracture surfaces frozen and the other relaxes them, a difference that has no cost on a termination breaking no bond. The maximum of $\sigma(d)$ in Equation~\ref{eq:cleavage} is an ideal cleavage stress of \SI{0.88}{\giga\pascal}, below the \SI{1.10 \pm 0.15}{\giga\pascal} measured for graphite \cite{Gould2013}.

\begin{table*}[tb]
\centering
\caption{Surface energy $\gamma$, Wulff area fraction, ionization potential IP, and electron affinity EA values of the four facets, the latter two referenced to the vacuum level.}
\label{tab:estabilidade}
\begin{tabular}{lcccc}
\toprule
Facet & $\gamma$ (J\,m$^{-2}$) & Area (\%) & IP (eV) & EA (eV) \\
\midrule
(100) & 0.323 & 17.1 & 5.36 & 4.15 \\
(010) & 0.869 &  4.4 & -   & -   \\
(001) & 0.097 & 71.4 & 5.77 & 4.95 \\
(210) & 0.643 &  7.0 & 5.81 & 4.56 \\
\bottomrule
\end{tabular}
\end{table*}

\begin{figure*}[tb]
\centering
\includegraphics[width=0.75\textwidth]{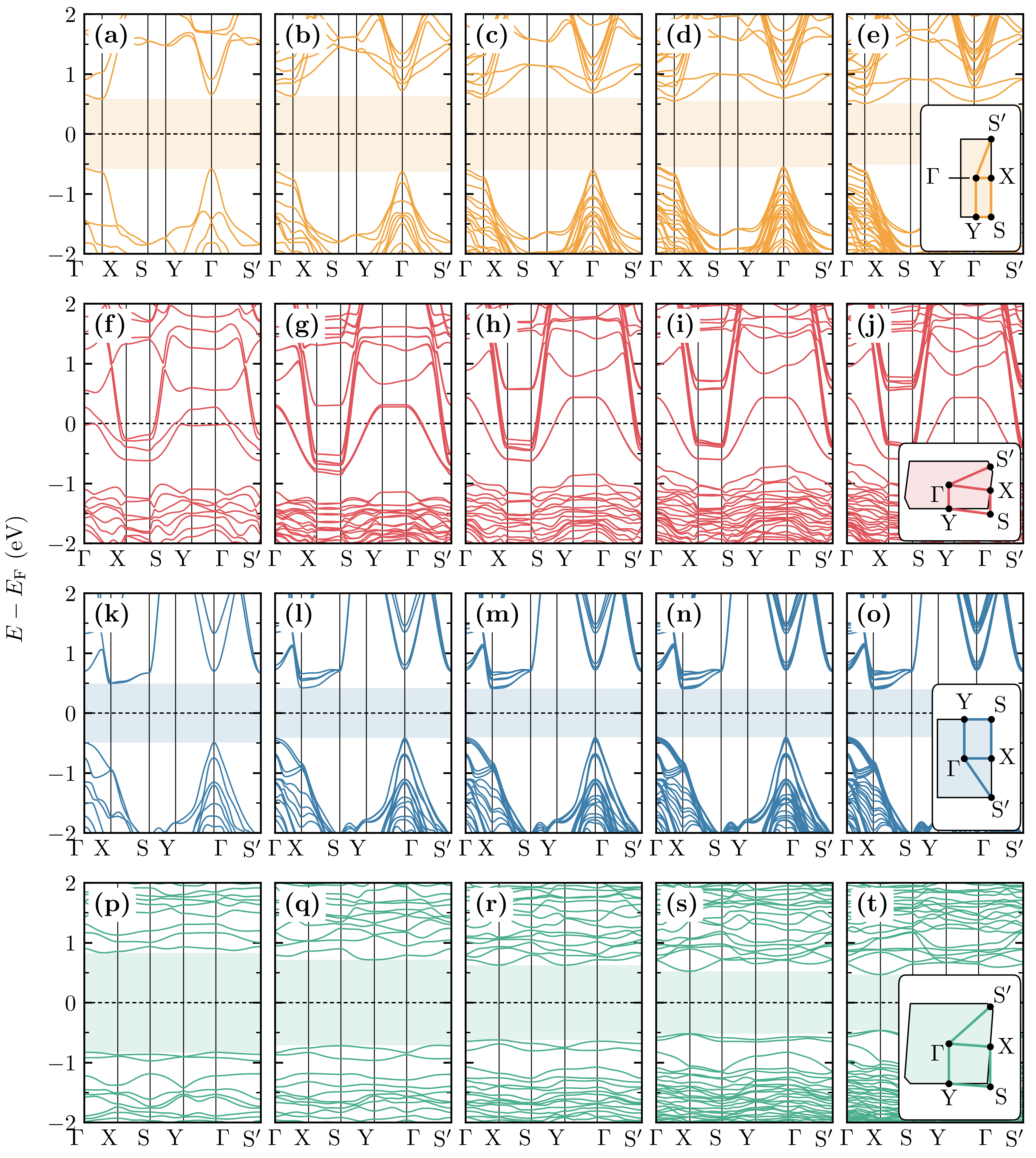}
\caption{Electronic band structures of the (100), (010), (001), and (210) surfaces, one facet per row, with slab thickness increasing from left to right. The inset shows the surface Brillouin zone and the band paths.}
\label{fig:bandas}
\end{figure*}

Having established which facets a ZrSe$_3$ crystal exposes, we now analyze the implications for its electronic properties. Figure~\ref{fig:bandas} collects the HSE06 band structures with SOC for the four facets across the thickness range, arranged with the facets in rows and the slab thickness in columns, and with the surface Brillouin zone and the band path shown as an inset. Because the slabs are periodic in two directions, the Brillouin zone is two-dimensional, and the thickness appears as the number of quantized subbands rather than as dispersion perpendicular to the surface, which every row shows as a progressive densification of the bands from left to right.

Across the four rows of Figure~\ref{fig:bandas}, three facets are semiconducting, and one is metallic. The (100) surface has an electronic band gap of \SI{1.20}{\electronvolt} at the slab thickness in Table~\ref{tab:eletronica}, the (010) surface is metallic at every thickness examined, (001) has \SI{0.82}{\electronvolt}, and (210) has \SI{1.25}{\electronvolt}. Every band gap is indirect. In the bulk crystal, the valence maximum lies at $\Gamma$, and the conduction minimum at $(0.5, 0, 0)$, the (100) and (001) surfaces preserve that arrangement in their surface zones, and the (210) surface separates its edges onto two different zone boundaries, at $(0.48, 0, 0)$ and $(0, 0.5, 0)$. The comparison with the optical measurements quoted above is therefore one of fundamental edges and not of the direct transitions those experiments also resolve. All three semiconducting values lie above the bulk band gap value of \SI{0.78}{\electronvolt} obtained here with the same functional, which in turn agrees with the \SI{0.75}{\electronvolt} previously reported for bulk ZrSe$_3$ with HSE06 \cite{Jin2015} and with the \SI{0.66}{\electronvolt} obtained at the $GW$ level \cite{Li2022}. The agreement across three independent routes to the bulk gap anchors the facet-resolved numbers that follow. The bulk entry of Table~\ref{tab:eletronica} carries no band-edge positions, since a three-dimensional periodic calculation has no vacuum region against which to reference them, and placing bulk levels on an absolute scale requires aligning the macroscopic average of the electrostatic potential between the crystal and a slab \cite{VandeWalle1987, Hinuma2014}. The thickest basal slab performs that role directly, since its interior is already bulk-like, and it places the valence-band maximum (VBM) at \SI{-5.77}{\electronvolt} and the conduction-band minimum (CBM) at \SI{-4.97}{\electronvolt} with a gap of \SI{0.80}{\electronvolt}, within \SI{0.02}{\electronvolt} of the bulk value. The bulk effective masses in the same row need no such reference, since a mass follows from the curvature of a band and is unchanged by a rigid shift of the energy scale, which is also why the metallic (010) surface has neither band edges nor band-edge masses.

Gaps are only part of what the exposed plane changes, since the band edges themselves move with it. Table~\ref{tab:eletronica} lists the VBM and the CBM values of each facet measured from the vacuum level, which are the IP and the EA of that facet. The IP varies between 5.36 and \SI{5.81}{\electronvolt} and the EA between 4.15 and \SI{4.95}{\electronvolt} across the three semiconducting facets. Hence, a single ZrSe$_3$ crystal presents band edges that differ by \SI{0.45}{\electronvolt} in the valence band and \SI{0.80}{\electronvolt} in the conduction band depending on which plane is exposed. Both spreads narrow as the slabs thicken, to 0.42 and \SI{0.62}{\electronvolt} at the largest thickness reached for each facet, so the facet ordering holds while the separation size does not. Shifts of this magnitude with surface orientation are known in other anisotropic semiconductors, with variations as large as \SI{0.9}{\electronvolt} reported across the orientations of orthorhombic SnS and attributed to the potential step set by the charge distribution normal to the surface \cite{Stevanovic2014}. For ZrSe$_3$, quantities usually treated as material constants, such as the barrier formed at a contact or the position of the band edges relative to a redox level, are facet properties rather than compound properties.

That same facet dependence also affects the effective masses of Table~\ref{tab:eletronica}. They are strongly anisotropic, as expected in a crystal built from chains, so each band edge is reported along the direction in which it disperses, which is the direction that carries transport. Each value comes from a dense sampling of reciprocal space centered on the band extremum, with a parabola fitted over an energy window of \SI{0.05}{\electronvolt}. Along the chain direction, the hole mass is \num{0.17} electron masses in the bulk crystal and the same on the basal plane and on (100), as a valence edge built from the chains should be. The basal electron mass of \num{0.21} is the one band-edge quantity on that plane that departs from the bulk, half again its value of \num{0.14}. The (100) surface increases the electron mass to \num{0.48}, and the (210) surface loses the light direction altogether, carrying holes of \num{2.39} and electrons of \num{1.96} electron masses, fourteen times the bulk values. Across the chains, the electron mass is 15.5 in the bulk crystal, and \num{14.2} on the basal plane, a hundredfold anisotropy between the two directions, which falls to \num{6.6} on (100) and to \num{1.6} on (210), so the one-dimensional character of the interior survives on the basal plane and disappears on the facet that carries the activity. The ordering across the three semiconducting facets follows how much each cut disturbs the chains: the basal termination leaves them intact, the (100) cut separates them laterally, and the (210) cut crosses them. Transport measurements on thin ZrSe$_3$ samples place the conduction predominantly on the outer Se atoms \cite{Hollmann2025}, consistent with a mobility that survives on the basal plane and degrades on the cut facets. The facet that the electrocatalytic measurements identify as active is therefore the one with the heaviest valence carriers, so its activity does not originate in carrier transport.

\begin{table*}[tb]
\centering
\caption{Electronic band gap, band edges referenced to the vacuum level, and effective masses in units of the free-electron mass. Each row comes from a single slab with thickness 15.2, 24.4, 25.2, and \SI{14.0}{\angstrom} for (100), (010), (001), and (210), respectively.}
\label{tab:eletronica}
\begin{tabular}{lcccccc}
\toprule
Facet & $E_{\mathrm{g}}^{\mathrm{PBE}}$ (eV) & $E_{\mathrm{g}}^{\mathrm{HSE}}$ (eV) & VBM (eV) & CBM (eV) & $m_{\mathrm{h}}^{*}$ & $m_{\mathrm{e}}^{*}$ \\
\midrule
(100) & 0.52 & 1.20 & $-5.36$ & $-4.15$ & 0.17 & 0.48 \\
(010) & 0.00 & 0.00 & -      & -      & -   & -   \\
(001) & 0.21 & 0.82 & $-5.77$ & $-4.95$ & 0.17 & 0.21 \\
(210) & 0.70 & 1.25 & $-5.81$ & $-4.56$ & 2.39 & 1.96 \\
Bulk  & 0.18 & 0.78 & -      & -      & 0.17 & 0.14 \\
\bottomrule
\end{tabular}
\end{table*}

\begin{figure}[!t]
\centering
\includegraphics[width=\linewidth]{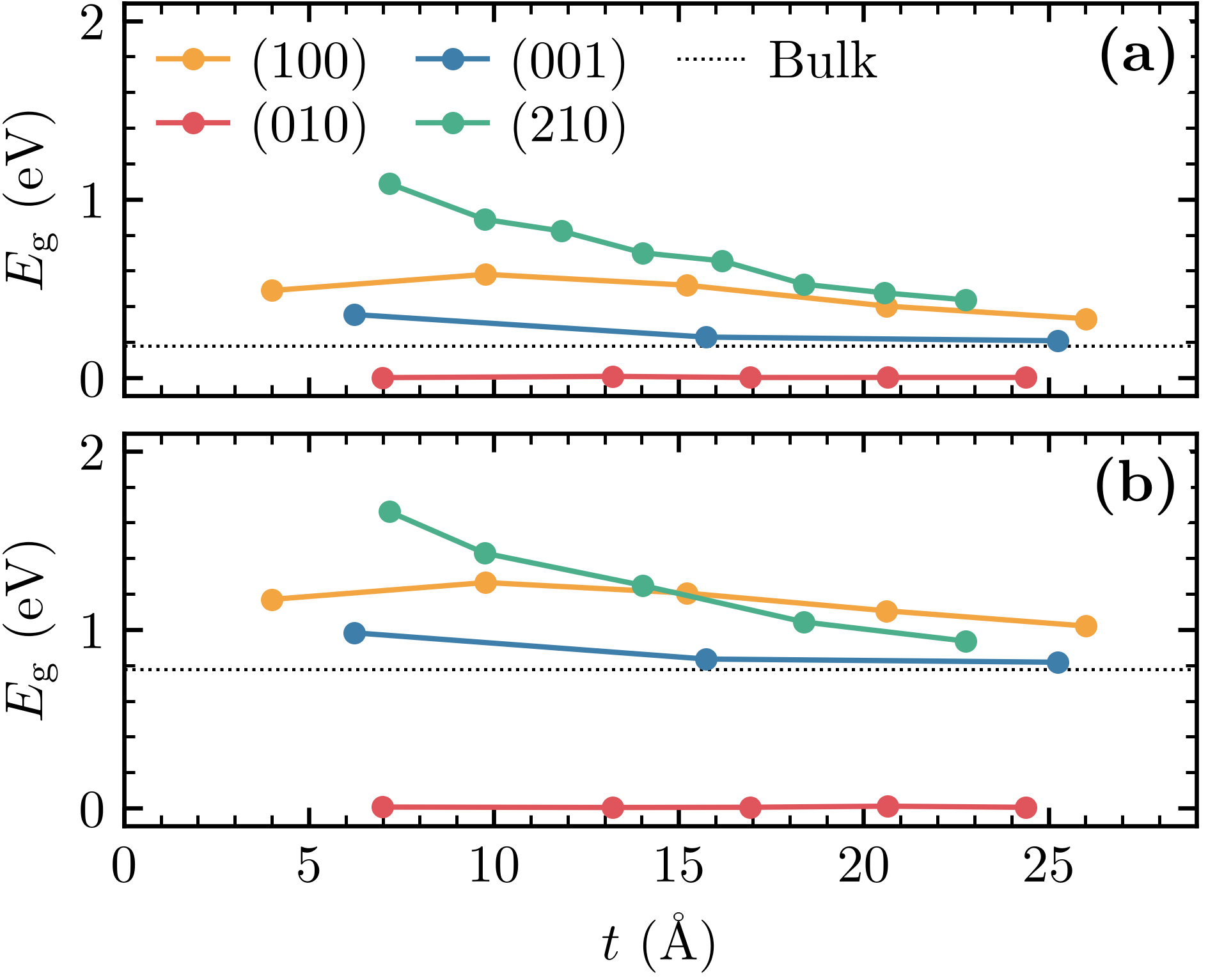}
\caption{Electronic band gap values as a function of slab thickness for the four facets and for the bulk crystal, with PBE (a) and with HSE06 including SOC (b).}
\label{fig:gap}
\end{figure}

All of those numbers were taken at one slab thickness per facet, and the crystals used in electrocatalytic and transport experiments are tens of nanometers thick, so which of those values apply to a real sample depends on how fast the surface electronic structure converges to the bulk limit. Figures~\ref{fig:gap}(a) and \ref{fig:gap}(b) show the band gap values of each facet against slab thickness with PBE and with HSE06 including SOC. For the semiconducting surfaces, the hybrid functional increases every band gap by 0.5 to \SI{0.7}{\electronvolt} relative to PBE, as expected from the self-interaction error that PBE carries in semiconductors, without changing the trend of any series, so what follows is indeed a property of the systems rather than of the functional.

The response to confinement sharply separates the facets. The (210) surface shows by far the strongest dependence, with its gap decreasing from 1.43 to \SI{0.94}{\electronvolt} between 10 and \SI{23}{\angstrom} and still decreasing at the largest thickness computed, while the basal plane varies by only \SI{0.18}{\electronvolt} from the monolayer to \SI{44}{\angstrom} and is within \SI{0.04}{\electronvolt} of the bulk value at \SI{25}{\angstrom}. The monolayer end of that series, \SI{0.98}{\electronvolt}, sits just under the 1.01 to \SI{1.17}{\electronvolt} range reported for the free-standing monolayer with hybrid functionals and without SOC \cite{Jin2015, Mortazavi2022, LiDaiZeng2015}. The \SI{0.06}{\electronvolt} by which SOC narrows the band gap here accounts for that difference. The (100) surface falls in between, and the (010) surface remains metallic throughout. A band gap that is nearly independent of thickness has also been found for TiS$_3$ and traced to band edges that sit on the central atoms of the layer and therefore do not experience the interlayer coupling \cite{KangWang2016}. The ZrSe$_3$ basal plane behaves in the same way, and the contrast with the (210) surface shows that the insensitivity is a property of the basal cut rather than of the compound.

Following the two band edges separately shows where the thickness dependence comes from. The VBM of the (001) surface moves by \SI{0.03}{\electronvolt} across the whole thickness range, while its corresponding CBM moves by \SI{0.15}{\electronvolt}, and the (100) surface behaves the same way, with \SI{0.10}{\electronvolt} against \SI{0.27}{\electronvolt}. The (210) surface follows the same pattern once its thinnest slab is set aside, with the valence edge moving \SI{0.11}{\electronvolt} against \SI{0.38}{\electronvolt} for the conduction edge between 10 and \SI{23}{\angstrom}. In contrast, the three-unit slab sits \SI{0.29}{\electronvolt} lower in the valence band than any thicker one. In device terms, the EA of ZrSe$_3$ is tunable through sample thickness, while the IP is not, which parallels the behavior reported for few-layer phosphorene, where the band edges and work function shift rapidly over the first few layers and then saturate \cite{Cai2014, Tran2014BP}. Measurements on ZrSe$_3$ show the transport gap widening as samples are thinned \cite{Thole2022, Hollmann2025}, in the same direction as the calculated trend. However, the transport gap and the Kohn-Sham gap are not the same quantity, and the comparison is one of trend rather than magnitude.

Taken together, surface energy, band-edge position, carrier mass, and thickness response place the four facets in the same order every time, pointing to a common structural origin. That origin is not the density of broken bonds but the fate of the Se-Se dimer. In the relaxed bulk crystal, every dimer measures \SI{2.379}{\angstrom}. After relaxation, the (100) and (001) slabs retain every dimer between 2.371 and \SI{2.379}{\angstrom} at all layers, including the outermost layer, whereas the (010) slab compresses its outermost dimers to \SI{2.326}{\angstrom} and stretches the pairs beneath them to \SI{2.632}{\angstrom}, and the (210) slab does the same at 2.352 and \SI{2.589}{\angstrom}, with a further set at \SI{2.435}{\angstrom} in between. The distortion is reproducible, since relaxations of the (010) slab started from independently perturbed configurations converge to the same set of bond lengths, and it persists across the (210) thickness series from three to nine structural units.

Projecting the density of states (DOS) onto individual Se atoms shows that these distorted pairs carry the frontier states. Integrating the projected DOS over the \SI{0.5}{\electronvolt} below the highest occupied level gives between 0.02 and 0.11 electrons per atom for dimers at the bulk length, irrespective of how close to the surface they lie, against 0.13 to 0.26 for the stretched pairs and 0.47 to 0.64 for the compressed ones. The (210) slab makes the point on its own, since it carries a compressed pair and an undistorted one at almost the same depth, and the compressed one holds fifteen times the weight of its neighbor. The band-edge weight therefore follows the dimer's deviation from its bulk length, not its proximity to the surface, and compression produces about four times the effect of stretching. This is the surface counterpart of the mechanism that organizes the family's bulk electronic structure, in which the position of the Se-Se $\sigma^{*}$ level relative to the Fermi energy determines whether a chain is metallic \cite{Canadell1990, CanadellWhangbo1991}.

Metallicity on the (010) surface follows the same rule, read at the Fermi level instead of at the valence edge. That cut distorts its pairs in both directions, compressing the outermost to \SI{2.326}{\angstrom} and stretching those below to 2.611 and \SI{2.632}{\angstrom}, and the states at the Fermi level sit on those pairs and on no others, with \num{0.79} states per electronvolt per atom on the compressed pair, 0.21 to \num{0.27} on the stretched ones, and under \num{0.02} on every pair left at the bulk length. The bulk precedent is NbSe$_3$, where the Se-Se separation sets which of the three inequivalent chains carry the metallic bands \cite{Canadell1990, Hodeau1978}, and the (010) cut reproduces that condition at a surface. Because the band edges coincide there, neither an IP nor an EA is defined for that facet, and the single quantity that characterizes it is a work function of \SI{5.78}{\electronvolt}. The metallicity stays on the surface, with the DOS at the Fermi level reaching \num{0.317} states per electronvolt per atom in the layers within \SI{3}{\angstrom} of either face against \num{0.078} in the interior, a ratio of \num{4.1} at HSE06 and \num{4.3} at PBE. Scanning tunneling microscopy on the ZrSe$_3$ basal surface has shown an unequal charge distribution over the three surface Se atoms \cite{Prodan2001}, and photoemission on the related ZrTe$_3$ resolves a Fermi-surface splitting attributed to a relaxation of the outermost layer that differs from the bulk \cite{Hoesch2009}, so surfaces in this family are known to depart electronically from their interiors.

Departing electronically from the interior does not create gap states. The band edges obtained from the surface layers and from the interior layers coincide to within \SI{0.08}{\electronvolt} on the (100), (001), and (210) surfaces, so no facet, including the two that distort their dimers, carries dangling-bond states in the gap. Since the (010) facet occupies only \SI{4.4}{\percent} of the equilibrium surface area, a ZrSe$_3$ crystal remains semiconducting overall while carrying a metallic minority facet, which is consistent with the semiconducting behavior measured on exfoliated material \cite{Thole2022}.

Those same band edges, read against an electrochemical scale, indicate the thermodynamic driving force for proton reduction. Transfer requires a conduction minimum above the hydrogen evolution level, which lies at \SI{-4.44}{\electronvolt} on the vacuum scale, and at zero bias only (100) qualifies, with its minimum \SI{0.29}{\electronvolt} above the level. The (210) and (001) minima sit 0.12 and \SI{0.51}{\electronvolt} below it and open only under cathodic bias, (210) first, since its conduction edge stays above the basal one at every thickness computed, by \SI{0.39}{\electronvolt} at the thicknesses of Table~\ref{tab:eletronica} and by \SI{0.10}{\electronvolt} at the thickest slabs. Ranked by electron availability alone, the three semiconducting facets go (100), (210), and then the basal plane, which comes last despite covering most of the crystal. That ranking does not match the microcell, where the activity followed (210) and not (100) \cite{CamposMata2026}. Electron availability therefore provides a thermodynamic criterion for proton reduction, but it does not by itself determine the activity. The one feature that distinguishes the active facet from the other two is the distorted Se-Se pair that (210) exposes and neither (100) nor (001) does, and we propose it as the origin of the activity measured there.

\section{Conclusions}

We have computed the values of surface energies, equilibrium morphology, and facet-resolved band-edge alignment of ZrSe$_3$ with a hybrid functional and SOC, covering the four crystallographic planes exposed by exfoliated crystals. The surface energies span nearly an order of magnitude, from \SI{0.097}{\joule\per\meter\squared} for the van der Waals basal termination to \SI{0.869}{\joule\per\meter\squared} for the (010) cut, which is perpendicular to the chains, and the Wulff shape built from them exposes the four observed planes, with the basal plane accounting for \SI{71.4}{\percent} of the area and the electrocatalytically active (210) facet for \SI{7.0}{\percent}. Cleaving the basal plane requires \SI{0.188}{\joule\per\meter\squared} and an ideal stress of \SI{0.88}{\giga\pascal}.

Electronically, the facets are not interchangeable. The ionization potential and the electron affinity shift by \SI{0.45}{\electronvolt} and \SI{0.80}{\electronvolt} across the three semiconducting facets. The carriers on (210) are more than an order of magnitude heavier than in the bulk, while the basal plane matches the bulk light hole, and the anisotropy between the two in-plane directions falls from a factor of a hundred in the interior to less than two on (210). The (010) facet is metallic at every thickness, and the crystal as a whole stays semiconducting because that facet occupies so little of the equilibrium area. Thickness primarily affects the conduction band, and far more strongly on (210) than on the basal plane.

A single structural variable accounts for the facet dependence. Cuts that preserve the Se-Se dimer at its bulk length of \SI{2.379}{\angstrom} yield surfaces electronically indistinguishable from the interior, while cuts that compress or stretch it concentrate the frontier states on the distorted pairs, with compression raising the band-edge weight about four times more than stretching. Because the same dimer controls whether a chain is metallic in the bulk compounds of the MX$_3$ family, the criterion carries over to ZrS$_3$, TiS$_3$, and HfS$_3$ as a testable expectation for which of their terminations will depart electronically from their bulk.

\section*{CRediT authorship contribution statement}
\noindent\textbf{Sandro G. Holanda}: Data Curation, Investigation, Formal analysis, Writing - original draft.
\textbf{Bruno Ipaves}: Investigation, Formal analysis, Writing - original draft.
\textbf{Astrid Campos-Mata}: Investigation, Writing - original draft.
\textbf{Shreyasi Chattopadhyay}: Investigation, Writing - review \& editing.
\textbf{Pulickel M. Ajayan}: Conceptualization, Resources, Supervision, Funding acquisition, Writing - review \& editing.
\textbf{Douglas S. Galv\~{a}o}: Conceptualization, Resources, Supervision, Funding acquisition, Writing - review \& editing.
\textbf{Marcelo L. Pereira Junior}: Conceptualization, Methodology, Software, Formal analysis, Investigation, Visualization, Supervision, Project administration, Writing - original draft, Writing - review \& editing.

\section*{Declaration of competing interest}
The authors declare no known competing financial interests or personal relationships that could have appeared to influence the work reported in this paper.

\section*{Data availability}
The data that support the findings of this study are available from the corresponding author upon reasonable request.

\section*{Acknowledgments}
All calculations reported here were performed on the supercomputer of the NanoEngineering Laboratory (NanoEng) at the University of Bras\'{i}lia, \url{https://nanoeng.unb.br}. A.C.-M. acknowledges funding from the Secretar\'{i}a de Ciencia, Humanidades, Tecnolog\'{i}a e Innovaci\'{o}n (SECIHTI) provided under the Ph.D. scholarship program (CVU 1051087). B.I. acknowledges financial support from the Brazilian National Council for Scientific and Technological Development (CNPq, process 153733/2024-1) and from the S\~{a}o Paulo Research Foundation (FAPESP, process 2024/11016-0). M.L.P.J. acknowledges financial support from the Federal District Research Support Foundation (FAPDF, grant 00193-00001807/2023-16), CNPq (grants 444921/2024-9 and 308222/2025-3), and the Coordination for the Improvement of Higher Education Personnel (CAPES, grant 88887.005164/2024-00). A portion of this project was supported by the U.S. Air Force Office of Scientific Research and Clarkson Aerospace Corp. under Award FA9550-24-1-0004. We also acknowledge support from INEO/CNPq and FAPESP grant 2025/27044-5.

\bibliographystyle{elsarticle-num}
{\raggedright
\bibliography{bibliography}
}

\end{document}